\documentclass[oldversion]{aa} 
\usepackage{graphicx}
\usepackage{aalongtable}
\usepackage{txfonts}
\usepackage{rotating}
\usepackage{lscape}
\newcommand{\gsim}{\;\lower.6ex\hbox{$\sim$}\kern-7.75pt\raise.65ex\hbox{$>$}\;}
\newcommand{\lsim}{\;\lower.6ex\hbox{$\sim$}\kern-7.75pt\raise.65ex\hbox{$<$}\;}

\begin{document}
\title{A sequence of nitrogen-rich very red giants in the globular
cluster NGC~1851\thanks{Based on observations collected at 
ESO telescopes under programme 188.B-3002}
 }

\author{
Eugenio Carretta\inst{1},
Valentina D'Orazi\inst{2,3},
Raffaele G. Gratton\inst{4},
\and
Sara Lucatello\inst{4}
}

\authorrunning{E. Carretta et al.}
\titlerunning{Nitrogen abundances in NGC~1851}

\offprints{E. Carretta, eugenio.carretta@oabo.inaf.it}

\institute{
INAF-Osservatorio Astronomico di Bologna, Via Ranzani 1, I-40127
 Bologna, Italy
\and
Dept. of Physics and Astronomy, Macquarie University, Sydney, 
NSW, 2109 Australia 
\and
Monash Centre for Astrophysics, School of Mathematical Sciences, Building 28,
Monash University, VIC 3800, Australia
\and
INAF-Osservatorio Astronomico di Padova, Vicolo dell'Osservatorio 5, I-35122
 Padova, Italy }

\date{}

\abstract{We present the abundances of N in a sample of 62 stars on the red
giant branch (RGB) in the
peculiar globular cluster NGC~1851. The values of [N/Fe] ratio were obtained by
comparing the flux measured in the observed spectra with that from synthetic
spectra for up to about 15 features of CN. This is the first time that N
abundances are obtained for such a large sample of RGB stars from 
medium-resolution spectroscopy in this cluster. With these abundances, we 
provide chemical tagging of the split red giant branch (RGB) found from 
several studies in NGC~1851. The secondary reddest sequence on the RGB is 
populated almost exclusively by N-rich stars, confirming our previous 
suggestion based on Str\"omgren magnitudes and colours. These giants are also,
on average,  enriched in $s-$process elements such as Ba, and are likely the 
results of pollution from low-mass stars that experienced episodes of third 
dredge-up in the asymptotic giant branch phase.
  }

\maketitle

\section{Introduction}
In the variegated landscape traced by multiple stellar populations in globular
clusters (GCs; see Gratton, Sneden \& Carretta 2004 and Gratton, Carretta \&
Bragaglia 2012a for recent reviews that summarize decades of studies on this
topic), some objects present a further level of complexity.

This is the case of the globular cluster NGC~1851. Less massive than GCs where
multiple main sequences were discovered for the first time (such as $\omega$
Cen, Anderson 1998; and NGC~2808, D'Antona et al. 2005), this cluster contained
evidence of multi-modality, starting with the bimodal distribution of stars on
its horizontal branch (HB; Walker 1992) and following with the finding of a
split subgiant branch (SGB). The origin of this feature is still debated, which
possibly invokes explanations of age or CNO-content  differences (Milone et al.
2008, 2009, Cassisi et al. 2008, Gratton et al. 2012b). Beside these features, a
split in the red giant branch (RGB) was recently detected by Han et al. (2009)
with $UVI$\ Johnson broadband photometry and by Lee et al. (2009a) using narrow
band Ca~II and Str\"omgren photometry. Side by side to a main RGB, another
sequence of  cluster red giant stars, less populated, is seen in these studies,
and it recently was shown to be connected to the faint SGB by Lardo et al.
(2012), using Str\"omgren photometry.

Three effects may produce a redder sequence of red giants in a stellar population:
\begin{enumerate}
\item The metal content (in particular the abundance of the $\alpha$-elements) 
is higher in the redder sequence. This implies an increase in the number of
free electrons, and as consequence of the opacity, giving a redder RGB.
\item The relative age of the two sequences is different with the reddest one
being younger, which causes the RGB to be redder and less steep.
\item Enhanced N, along with C, may result in a redder sequence.
\end{enumerate}

Lee et al (2009b) inserted NGC~1851 in the sample used to devise an enrichment
scenario for multiple populations in GCs where supernovae (SNe) also do
contribute to the pollution of the intracluster gas. This scenario is ruled out
in most clusters, due to the constancy of their $\alpha$-element content
(Carretta et al. 2010a), but we recently found a contribution by core-collapse
SNe in a fraction of RGB stars (Carretta et al. 2011) from the abundance
analysis of Ca and other $\alpha$-elements in NGC~1851. Moreover, a spread in
[Fe/H]\footnote{We adopt the usual spectroscopic notation, which is for any given
species X, [X]=  $\log{\epsilon(X)_{\rm star}} - \log{\epsilon(X)_\odot}$ and  
$\log{\epsilon(X)}=\log{(N_{\rm X}/N_{\rm H})}+12.0$\ for absolute number 
density abundances.}, which was small but exceeding the observational errors,
was detected in the same analysis, confirming earlier results by Yong and
Grundahl (2008) and Carretta et al. (2010b). Metal-rich and metal-poor giants 
also show a different radial distribution with more metal-rich stars being less
centrally concentrated (see Figure 6 in Carretta et al. 2011); however, these
stars are not segregated exclusively along the reddest RGB sequence.

A spread in the total CNO sum was found
by Yong et al. (2009) in a limited sample of four RGB stars in NGC~1851.
Enhanced CNO abundances in a fraction of stars were also proposed by Cassisi et
al. (2008) as a possible explanation for the split SGB. The implication is that
small mass AGB stars (those experiencing episodes of third dredge-up) also
contribute to the budget of nuclearly processed matter to construct a second
generation of stars in NGC~1851. Thus, a fraction of stars may be found to be
younger than the other. However, Villanova et al. (2010) challenged this
hypothesis in part by finding that the RGB stars on the two sequences do not show any
significant difference in the total C+N+O sum but only in their content of
$s-$process elements, in particular, for Ba. On the other hand, variations of
$s-$process elements like Zr and La were found to be correlated to light
elements (like Al) by Yong and Grundahl (2008): this again calls for
contribution from thermally pulsing AGB stars of low initial masses.

Using a large set of elemental abundances for a large sample of giants, we
recently proposed a classification scheme that divides the RGB stars in NGC~1851
into two populations using a combination of Fe and Ba (Carretta et al. 2011).
Our data show that stars are nicely segregated along the split RGB according to
their Ba abundances with Ba-poor stars located mostly (but not only) on the
bluest side of the RGB, whereas only Ba-rich stars are found on the less
populated reddest sequence. This result puts on a more statistically robust base 
similar findings by Villanova et al. (2010).
We recently find evidence of a statistically significant correlation between Al
and Ba abundances in the more metal-rich component of RGB stars (Carretta et al.
2012a).

Moreover, using the comparison of Str\"omgren colours and synthetic spectra
computed with varying C and N abundances, we demonstrated that stars on the
reddest sequence are very good candidates to be C-rich and N-rich, while stars
that are C-rich and N-poor (i.e. O-poor) likely lie side-by-side on the bluest
sequence with N-poor and C-normal stars (see figures 23 and 28 in Carretta et
al. 2011). 

The existence of a Na-O anticorrelation in each population suggests that the
classical chain of events (first generation, pollution, second generation,
marking the birth of a genuine GC, as proposed in Carretta et al. 2010c) was
present in each component. In turn, this may suggest a merger of two formerly
individual clusters into an ancestral galaxy or two different, yet spatially 
close regions of star formation.

Villanova et al. (2010) found a different N content in the two group of stars
with different Ba abundances. Campbell et al. (2012) recently studied the CN
bandstrengths of a sample of 21 AGB stars and 17 RGB stars in NGC~1851; however,
they were selected
on the $V,B-V$ color-magnitude diagram, where the red and blue RGB
sequence cannot be separated. In their two samples they found a quadrimodal
distribution of CN strengths with the more CN-weak populations coupled with low
Ba abundances and the more CN-strong peaks of the distributions associated with
high Ba values.

Apart from these limited samples of RGB stars, no extensive
spectroscopic analysis of the N abundance of objects on the reddest sequence was
made up to now. With the present paper, we intend to fill this gap, deriving N
estimates for about 60 red giants and providing another constraint for a better
understanding of the complex star formation in NGC~1851.

\section{Observations and analysis}

The same observational data was recently used by us (Carretta et al. 2012a)
to derive Al abundances for a large sample of RGB stars in NGC~1851: two
exposures of 600 sec each made with FLAMES-GIRAFFE mounted at VLT-UT2 and the
high-resolution grating HR21 (R=17,300, spectral range from 8484~\AA\ to
9001~\AA). We refer to Carretta et al. (2012a) for details on the treatment of
spectra (flat-fielding, calibration in wavelength, and sky subtraction). As before,
atmospheric parameters (as well as abundances of Fe, O, Na, Mg, Ba etc.) were 
adopted from the highly homogeneous analysis by Carretta et al. (2011).
The present sample is restricted to a magnitude range between $V\sim 17$ and 
$V\sim 15$.

In Carretta et al. (2012a), only a single feature was used to
derive the N abundance, which, assuming [C/Fe]=0, reproduces the CN band strength 
to account for a possible contamination of Al lines.
However, N lines are ubiquitous in the observed spectral range, and instead of
treating them purely as contaminant, we can use them to derive N
abundances for a large number of giants in NGC~1851.
Our approach closely follows the one adopted in Carretta et al. (2012b) for the
globular cluster M~4, which has a metallicity [Fe/H]$=-1.20$ dex (Carretta et al.
2009) similar to that of NGC~1851 ([Fe/H]$=-1.16$ dex). However, we remind the
reader that
the proprietary spectra taken in the case of M~4 were acquired to maximise the
signal around the Al lines, whereas the spectra with HR21 retrieved from the
public ESO archive for NGC~1851 were observed in a public survey designed for
other purposes and do not have an optimal S/N ratio for the task of abundance
analysis for elements like N. 

To deal with this issue, we used an improved version of the procedure adopted 
in Carretta et al. (2012a). Briefly, 18 spectral regions dominated by CN bands
and 8 regions that are apparently line-free were selected on a master spectrum obtained
by summing up the ten spectra of giants with the best S/N value. The list of
these regions is given in Carretta et al. (2012b).
For each CN feature, we measured the average flux within the in-line region, and
this was normalized using a weighted reference continuum where the weights are
equal to the width of the continuum regions.
The N abundance from each feature was then obtained by comparing the normalized
flux with the fluxes measured in the same way on three synthetic spectra 
computed using the Kurucz (1993) grid of model atmospheres (with the
overshooting option switched on). The  line list is described in Carretta et al.
(2012a), and the synthetic spectra were computed with the package ROSA (Gratton
1988) by assuming  [C/Fe]=0.0 dex, three evenly spaced values of nitrogen,
[N/Fe]=-0.5, +0.0, and +0.5 dex, and the appropriate atmospheric parameters for
each star.

Because of the non-optimal quality of the spectra,  the final abundances of N
were obtained after
applying a severe k$\sigma$-clipping at 1.5 $\sigma$ to the average abundance
from individual features in each star. On average, only ten features survived
this culling process. The average abundance ratio we found is 
[N/Fe]$=-0.082 \pm 0.037$ dex with rms=0.294 dex from 62 stars.

As explained in Carretta et al. (2012a), an error due to flux measurement can be
estimated for abundances derived through this procedure. This is obtained by
estimating photometric errors from the S/N ratio of the spectra and from the
width within each of the reference continuum and in-line regions  (see details
in Carretta et al. 2012a). The N abundances, derived by summing the original value
of the CN band strength and its error, allow us to estimate the error associated to
N abundances due to errors in the flux measurements. When we applied this
procedure to a random CN feature used to estimate the N abundance, we found an
average error of $+0.120 \pm 0.028$ dex from 62 stars in NGC~1851, which is a
conservative estimate since the final N abundances rest on approximatively ten
features per star. Actual star-to-star errors (those really relevant in the
problem under scrutiny here) are then about 0.04 dex. Internal errors imputable
to errors in the adopted atmospheric parameters have a similar impact, since
star-to-star errors in effective temperature, gravity, model abundance and
microturbulent velocity are only 4 K, 0.041 dex, 0.034 dex, and 0.16 km/s,
respectively, as estimated in Carretta et al. (2011). The total error expected
in N abundance due to these uncertainties is 0.137 dex. However, these
uncertainties have a negligible impact on our results, since they affect stars
on both sequences.

Unfortunately, we have no stars in common with the small sample studied by
Yong et al. (2009). In our sample, there are two stars in common with the
analysis of Villanova et al. (2010), one in their Ba-poor group (star 35999)
and the other in their Ba-rich group (star 29203). Taking into account the C
abundance (assumed, in our case; derived, in Villanova et al. 2010), we found a
difference (in the sense Villanova minus us) of 0.30 dex in [N/Fe] for
star 35999 and only 0.10 dex for star 29203. The comparison is not conclusive,
owing to the different features and methods adopted in the two studies and resting
on two objects only. For the aim of this analysis, the comparison of relative N
abundances along the two sequences of the RGB, it is, however, the internal
self-consistency that actually matters.

\section{Results and discussion}

\begin{figure}
\centering
\includegraphics[scale=0.40]{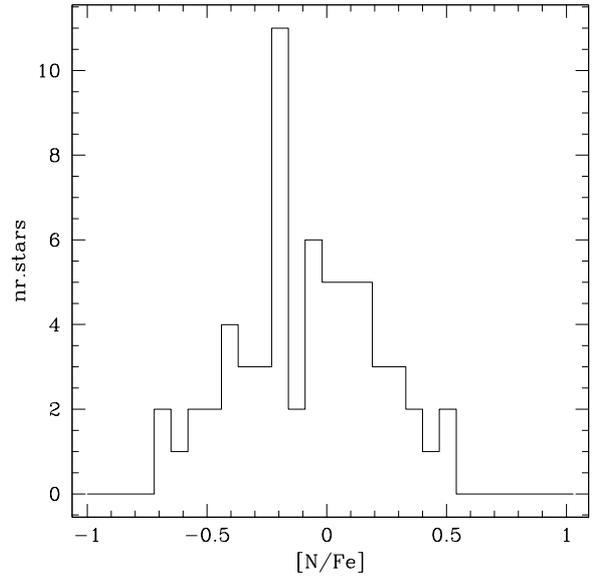}
\caption{Distribution of the [N/Fe] ratios obtained among RGB stars in 
NGC~1851.}
\label{f:histn18}
\end{figure}

The derived values of the [N/Fe] ratios for the 62 stars in our sample are
listed in Table 1, and their distribution is shown in Fig.~\ref{f:histn18}.

As a working hypothesis, we defined the giants having [N/Fe]$\leq -0.15$ dex as
N-poor stars  and those with [N/Fe]$>-0.15$ dex as N-rich giants . The
separating value simply corresponds to the dip in the histogram in
Fig.~\ref{f:histn18}.

\begin{figure}
\centering
\includegraphics[scale=0.40]{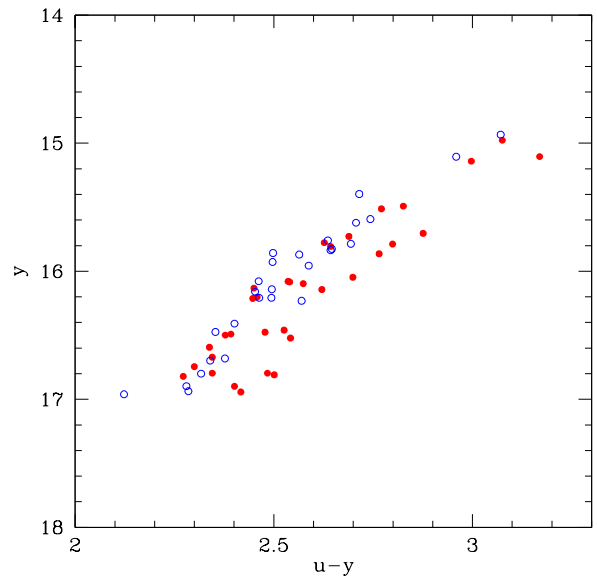}
\caption{Str\"omgren $y,u-y$ colour-magnitude diagram of RGB stars in our sample
in NGC~1851. Blue open circles are for giants with [N/Fe]$\leq -0.15$ dex and
filled red circles are for stars with [N/Fe]$>-0.15$ dex.}
\label{f:yuy}
\end{figure}

These two groups are indicated with different symbols in Fig.~\ref{f:yuy}, where
stars in our sample are plotted in the $y,u-y$ colour-magnitude diagram (CMD)
using the unpublished Str\"omgren photometry kindly provided to us by 
Jae-Woo Lee (private communication). From this CMD it is clear that all stars 
on the less populated, redder sequence are N-rich apart from a single outlier
(possibly due to a combination of uncertainties in the photometry and/or N
abundance). The bulk of giants, however, lie on the bluest RGB, which consists 
of a mix of N-rich and N-poor stars.

\begin{figure}
\centering
\includegraphics[bb=19 146 364 716, clip, scale=0.52]{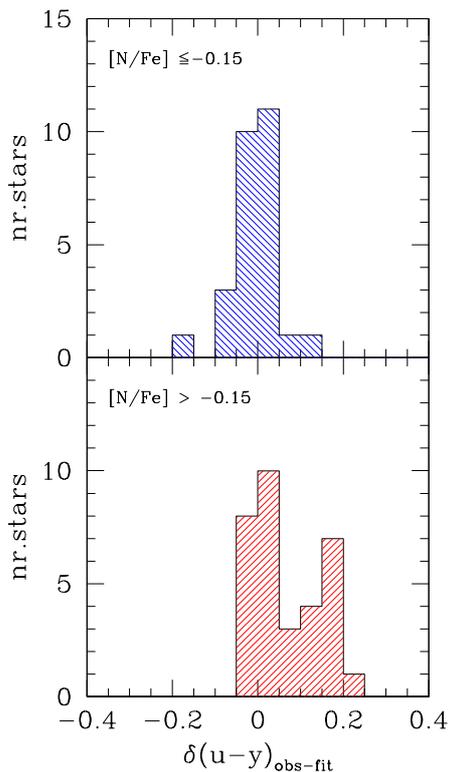}
\caption{Distribution of the differences $\delta(u-y)$ between the observed 
$u-y$ colour index of each star and the $u-y$ colour of the mean ridge line
through the  N-poor stars at the same $y$ magnitude. The distribution is
plotted separately for N-poor giants and N-rich stars (upper and lower panels,
respectively).}
\label{f:lmh2}
\end{figure}

To represent this finding in a more quantitative way, we traced by eye the  mean
ridge line passing through the bluest sequence in the  $y,u-y$ plane, using  the
N-poor stars as reference. Then, for each star, we computed the colour
differences between the $u-y$ observed colour and the colour of the mean ridge
line at the same magnitude $y$. The distributions of these residuals are plotted
in Fig.~\ref{f:lmh2} separately for the groups of N-rich (lower panel) and
N-poor (upper panel) giants.

\begin{figure}
\centering
\includegraphics[scale=0.40]{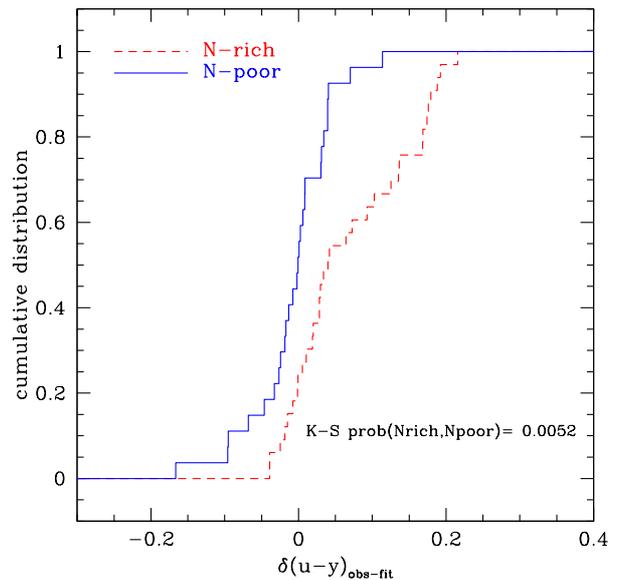}
\caption{Cumulative distributions of the residuals in $u-y$ colours between each
star and the colour of the mean ridge line through N-poor stars. Solid line is
for the distribution of N-poor giants and dashed line for the N-rich stars. The 
probability for the two distributions of being extracted from the same parent
sample under the Kolmogorov-Smirnov statistical test is given.}
\label{f:distrduy}
\end{figure}

The two distributions are clearly different, as confirmed by a statistical
Kolmogorov-Smirnov test (see Fig.~\ref{f:distrduy}): the two populations have a
very low probability (0.0052) of being extracted from the same parent sample.
Thus, the main conclusion of the present study is that the redder, less
conspicous, secondary sequence on the RGB of NGC~1851 seems to be populated only
by N-rich stars. The N-poor stars are almost completely segregated on the bluest
part of the RGB with a fraction of N-rich giants.

There are six stars in common between our analysis and the sample of 17 RGB stars
analyzed by Campbell et al. (2012). Using the Str\"omgren photometry available
to us for these 6 objects, we see that the two stars with the highest values of
$\delta$S(3839) (stars 41113 and 32112 in Table 1 of Campbell et al. 2010) lie
on the redder RGB sequence. The other four stars (46228,44414,47385, and 44939)
are located on the blue sequence and all have lower values of $\delta$S(3839).

In our study, we presented the first spectroscopic derivation of N abundances of
giants in this cluster made for such a large number of stars on the RGB, which 
nicely supports the results by Villanova et al. (2010) and Campbell et al.
(2012) based on much more limited samples. While we cannot state anything on
the total C+N+O sum with not knowing the C abundance of the individual objects,
we are able to spectroscopically ascertain  that stars are located on the redder
sequence in NGC~1851 because they  are N-rich.

Moreover, the present work strongly confirms the findings by Carretta et al.
(2011) based on an independent approach. Simply from Str\"omgren photometry and
explorative spectral synthesis of a C-rich/N-rich and of a C-rich/N-poor typical
giant in NGC~1851 from Carretta et al. (2011), we concluded that stars on the
redder sequence were ``good candidate to be C-rich and N-rich (hence O-poor). On
the other hand, the C-rich and N-poor stars (i.e., O-rich) may not be separated
from the other N-poor and C-normal stars in none of the Str\"omgren indices"
(Carretta et al. 2011). 
{\it In the present paper, we were able to confirm this statement by deriving 
directly the N abundances of 62 RGB stars.}

For 60 stars in our sample we have the values of the $hk$ index, kindly
provided by Jae-Woo Lee (private communication). The mean value of the index for stars on the redder
sequence is +0.927 ($rms=0.104$), whereas we found a mean value of +0.799
($rms=0.114$) for stars with normal N abundances. This difference can be 
compared to the Full Width Half-Maximum (FWHM) of the RGB in NGC~1851 in $hk$ found by Lee et al. (2009b):
0.182 with a measurement error of 0.013 (their supplementary Table 3). A direct
comparison to the N abundances found for BHB stars by Gratton et  al. (2012c)
is not easy because different abundance indicators (high-excitation lines) were
used in that study, and non-LTE corrections are likely not negligible. Moreover,
this comparison is hampered by the lack of knowledge of the absolute C values
for our sample, where N abundances are derived from CN features. Finally, we
found  no evidence of radial variations in N abundances in the present sample.

Recently, Lardo et al. (2012) showed that the red sequence on the RGB of
NGC~1851 is photometrically connected to the faint SGB discovered by Milone et
al. (2008). This result, when coupled with evidence from spectroscopy in the
present and previous studies (e.g. see Yong et al. 2009 and Villanova et al.
2010 for a different view), gives support to the suggestion by Cassisi et al.
(2008) that an increased amount of CNO (a factor about two) in stars of the
faint SGB may explain the difference between the split sequences.

It is also well assessed (Villanova et al. 2010, Carretta et al. 2011, Campbell
et al. 2012) that only Ba-rich stars are found on the redder sequence of the
RGB with Ba-poor stars mostly (but not exclusively) confined to the more
populated bluest part of the giant branch. This close similarity with the
pattern found here for N abundances points towards the action of first generation
polluters which experience episodes of third dredge-up, stars with initial mass in
the range 1.5-3 $M_\odot$.
Moreover, Gratton et al. (2012b) investigated the abundance
pattern of stars belonging to the two SGBs in NGC~1851; they found that stars on
the faint SGB are Sr and Ba-rich, while the group on the bright SGB show a low
content of these elements. These findings point out that the observed spread of
$s-$process elements, currently traced from the SGB up to the RGB, is real in
this cluster.

On the other hand, the correlation between proton-capture elements (like Na 
and Al)
and the amount of $s-$process elements found by us (Carretta et al. 2011,
2012a) and other studies (Yong and Grundahl 2008; Villanova et al. 2010)
presents a problem related to the different evolutionary times of different
classes of first generation polluters.
As discussed by Carretta et al. (2011), if the polluters were AGB stars, those
producing proton-capture elements, such as Al, are in the mass range between 4 and
8 $M_\odot$. This leaves a time delay of about 200 Myr before the effects of
mass loss from low-mass AGB stars (producing the bulk of $s-$process and
enhanced CNO sum) may be incorporated in the gas forming stars of the second
generation.

A possible solution of this puzzle is if the two sources of pollution are
decoupled in origin. In other words, suppose that the gas enriched in $s-$process
elements by low-mass AGB stars is not forged by stars of the forming
proto-cluster. This gas may, for example, be collected in the potential well of
a star-forming region immersed into a larger dwarf galaxy where matter 
leftover by a generation of low-mass AGB stars slowly inflows towards the
denser proto-cluster regions. Here, this gas ``unprocessed" by first generation
cluster stars may combine with nuclearly processed ejecta and eventually form
stars of a second generation that are rich in both Al and N and $s-$process elements.
Such a variant to the first-generation-pollution-second generation sequence,
where part of the pollution is not due to the same cluster stars, was originally
introduced by Bekki et al. (2007).

As an alternative, if fast-rotating massive stars (Decressin et al. 2007) are
responsible for the enrichment in proton-capture elements like Na and Al,
then the low-mass AGB stars within the first generation of cluster stars may
contribute to the enrichment without any particular problem that is related to the age
difference between intermediate and low-mass AGB stars.

\begin{acknowledgements}
VD is an ARC Super Science Fellow. This work was partially funded by the PRIN
INAF 2011 grant ``Multiple  populations in globular clusters: their role in the
Galaxy assembly" (PI E. Carretta), and the PRIN MIUR 2010-2011, ``The Chemical
and Dynamical Evolution of the Milky Way and Local Group Galaxies'', (PI F.
Matteucci), prot. 2010LY5N2T.  We thank  \v{S}ar\={u}nas Mikolaitis for sharing
with us the line list for CN provided  by B. Plez, prof. Jae-Woo Lee for sharing
his unpublished photometry, and Sandro Villanova for sending us C, N, O
abundances and coordinates for individual stars of its sample. This research
has made use of the SIMBAD database (in particular  Vizier), operated at CDS,
Strasbourg, France and of NASA's Astrophysical Data System.
\end{acknowledgements}

\clearpage
\begin{table*}
\centering
\caption[]{Abundance ratios [N/Fe] for RGB stars in NGC~1851.}
\begin{tabular}{lccrlccr}
\hline
star  & [N/Fe]  & rms  & nr$^1$ & star & [N/Fe]  & rms  & nr$^1$ \\
      & (dex)   & (dex)&     &      & (dex)   &	(dex)&    \\
\hline        
  13618 & $-$0.223 &0.165 &14 &  32256 & $-$0.195 &0.134 &11 \\
  14827 &   +0.032 &0.424 &10 &  35750 & $-$0.033 &0.215 &10 \\
  16120 & $-$0.214 &0.106 &11 &  35999 & $-$0.031 &0.125 & 7 \\
  20189 & $-$0.064 &0.231 & 9 &  36292 &   +0.508 &0.415 & 8 \\
  20426 & $-$0.473 &0.250 &11 &  38484 & $-$0.201 &0.236 &14 \\
  20653 &   +0.227 &0.119 & 9 &  38818 &   +0.026 &0.222 & 8 \\
  20922 & $-$0.170 &0.025 & 8 &  39364 & $-$0.204 &0.255 &16 \\
  21453 &   +0.052 &0.040 & 8 &  40300 & $-$0.106 &0.084 &11 \\
  21830 &   +0.086 &0.280 &11 &  40615 & $-$0.035 &0.033 & 7 \\
  22360 & $-$0.146 &0.243 &11 &  41113 & $-$0.054 &0.076 & 8 \\
  22588 & $-$0.417 &0.060 & 7 &  41855 & $-$0.387 &0.500 & 9 \\
  22813 & $-$0.310 &0.158 &12 &  43528 &   +0.389 &0.397 &11 \\
  23647 & $-$0.078 &0.109 &10 &  44224 &   +0.092 &0.665 &12 \\
  23765 &   +0.161 &0.081 &11 &  44414 & $-$0.196 &0.085 & 9 \\
  25037 & $-$0.265 &0.409 &13 &  44939 & $-$0.264 &0.060 &11 \\
  25497 &   +0.173 &0.190 & 8 &  45006 & $-$0.575 &0.418 &13 \\
  25678 & $-$0.709 &0.699 &11 &  45090 & $-$0.197 &0.256 &13 \\
  25799 & $-$0.570 &0.278 &10 &  45413 &   +0.300 &0.172 &11 \\
  26532 &   +0.178 &0.135 & 7 &  46228 & $-$0.195 &0.146 &12 \\
  26552 &   +0.035 &0.123 &10 &  46657 &   +0.063 &0.449 &12 \\
  26880 &   +0.424 &0.344 & 8 &  46958 & $-$0.462 &0.124 &10 \\
  27491 & $-$0.402 &0.089 & 7 &  47385 & $-$0.428 &0.127 &10 \\
  28116 &   +0.195 &0.068 & 7 &  47795 &   +0.380 &0.198 &13 \\
  29203 &   +0.313 &0.161 &13 &  48085 & $-$0.248 &0.037 &10 \\
  29470 &   +0.493 &0.124 & 9 &  48277 &   +0.212 &0.238 & 8 \\
  30286 &   +0.150 &0.116 & 8 &  48388 &   +0.063 &0.261 &14 \\
  31284 & $-$0.186 &0.116 &11 &  49965 &   +0.177 &0.133 & 8 \\
  31399 &   +0.317 &0.247 & 9 &  50876 & $-$0.670 &0.132 & 8 \\
  31463 & $-$0.301 &0.390 &10 &  50973 & $-$0.330 &0.071 & 9 \\
  31520 & $-$0.002 &0.162 & 9 &  51311 & $-$0.225 &0.207 &10 \\
  32112 &   +0.049 &0.137 &13 &  52570 & $-$0.639 &0.700 &11 \\
\hline
\end{tabular}
\label{t:abu}
\begin{list}{}{}
\item[1-] number of CN features used to derive [N/Fe]
\end{list}
\end{table*}

\end{document}